\documentclass[a4paper,reqno]{amsart}

\usepackage{amsmath, amssymb, amsthm, eucal}
\usepackage{graphicx}
\usepackage[shortlabels]{enumitem}

\usepackage{xr-hyper}
\usepackage[colorlinks=true]{hyperref}
\usepackage{verbatim}
\usepackage{pdfsync}

\usepackage{comment}

\usepackage{tikz}
\usepackage{scalefnt}

\usepackage[textheight=630pt,
  textwidth=468pt,
  centering]{geometry}

\usepackage{color}

\newtheorem{theorem}{Theorem}

\theoremstyle{definition}
\newtheorem{definition}[theorem]{Definition}
\newtheorem{remark}[theorem]{Remark}

\newcommand{\norm}[1]{\Vert #1 \Vert}

\newcommand{\RR}{\mathbb{R}}

\newcommand{\CC}{\mathbb{C}}

\newcommand{\nA}{\mathcal{A}}

\DeclareMathOperator{\tr}{tr}

\DeclareMathOperator{\id}{id}

\begin{document}

\title{Low-rank matrix recovery via rank one tight frame measurements}
\date{\today}
\author{Holger Rauhut}
\address{Lehrstuhl C f{\"u}r Mathematik (Analysis), RWTH Aachen University, Pontdriesch 10, 52062 Aachen, Germany}
\curraddr{}
\email{rauhut@mathc.rwth-aachen.de}
\thanks{}

\author{Ulrich Terstiege}
\address{Lehrstuhl C f{\"u}r Mathematik (Analysis), RWTH Aachen University, Pontdriesch 10, 52062 Aachen, Germany}
\curraddr{}
\email{terstiege@mathc.rwth-aachen.de}
\maketitle

\begin{center}
\small


\end{center}

\begin{abstract}
The task of reconstructing a low rank matrix from incomplete linear measurements arises in areas such as machine learning, quantum state tomography
and in the phase retrieval problem. In this note, we study the particular setup that the measurements are taken with respect to rank one matrices
constructed from the elements of a random tight frame. We consider a convex optimization approach and 
show both robustness of the reconstruction with respect to noise on the measurements as well as stability with respect to passing to approximately
low rank matrices. This is achieved by establishing a version of the null space property of the corresponding measurement map.
\end{abstract}


\medskip


\section{Introduction}
Compressed sensing \cite{fora13} predicts that sparse vectors can be reconstructed stably from an incomplete 
and possibly noisy set of linear (random) measurements via efficient algorithms including $\ell_1$-minimization. This theory has been extended to the reconstruction of low rank matrices from incomplete measurements. 	Initial contributions \cite{fapare10} analyzed Gaussian random measurement maps
having no structure at all. However, in quantum state tomography, for instance, one does require structure. 
A particular setup considers measurements with respect to rank-one matrices which makes the analysis more difficult because
of a reduced amount of (stochastic) independence. In \cite{KRT} and \cite{KKRT}, such rank one measurements
consisting of projections on random vectors drawn from a Gaussian distribution or from a complex projective $4$-design  are considered. In this note, we extend this to the case that the measurements project onto the elements from a random  tight frame. In order to make this more precise we recall the setup of \cite{KKRT}.


We consider rank one measurements of an (approximately) low-rank Hermitian matrix $X\in\mathcal H_n$  of the form
$ \nA(X)$, where the linear measurement map $\nA$ is given as
\begin{equation}\label{eq:MeasurementProcess}
\nA:\mathcal H_n\to\RR^m,\quad Z\mapsto\sum_{j=1}^m\tr(Za_ja_j^*)e_j.
\end{equation}
Here, $\mathcal H_n$ denotes the space of $n\times n$ complex Hermitian matrices, $e_1,\ldots,e_m$ denote the standard basis vectors in $\RR^m$ and $a_1,\ldots,a_m\in\RR^{n}$ are  measurement vectors. 
Taking into consideration the  presence of noise we write
\begin{equation}
\label{eq:measurements}
b=\nA(X)+w,
\end{equation}
where $w\in\RR^m$ is a vector of perturbations. 
We consider the following noise constrained nuclear norm minimization problem
\begin{equation}\label{eqNNMinimization}
\underset{Z\in\mathcal H_n}\min\norm{Z}_1\quad\mbox{subject to}\;\norm{\nA(Z)-b}_{\ell_q}\leq\eta,
\end{equation}
where $\eta$ denotes a known estimate of the noise level, i.e., $\|w\|_{\ell_q} \leq \eta$ for some $q\geq 1$ (including the case $q=\infty$). 
Here and in the sequel, we denote by $\norm{Z}_1$ the nuclear norm of $Z$ (the sum of its singular values). Similarly, $\norm{Z}_2$ denotes the Frobenius norm of $Z$ (the $\ell_2$-norm of the vector of singular values of $Z$), and $\norm{Z}_{\infty}$ denotes the maximal singular value of $Z$.

We are interested in choosing a minimal number $m$ of measurements (ideally smaller than $n^2$) that still allows reconstruction of $X$ of (approximately)
rank $r$ from $b = \nA(X)+w$. In this note, we choose the measurement vectors $a_1,\hdots, a_m$ from a tight frame, that is, the matrix $M \in \RR^{m \times n}$ whose rows are the vectors $a_j$ satisfies $M^*M = \id$. In order to analyze this setup rigorously, we introduce randomness. To this end we adopt the notation of \cite{TroppComparison}, and denote by  $\mathbb{V}_n^m$ the Stiefel manifold  consisting of the real valued $m\times n$ matrices $M$ with  the property $M^*M=\id$. 

\begin{definition}
Let  $m\geq n$. A  $m\times n$ {\emph random tight frame} is the set of rows of a $m\times n$ matrix with orthonormal columns which is drawn uniformly from the Stiefel manifold $\mathbb{V}_n^m$. 
\end{definition}

We also use the following notation. Given a complex valued  $n_1\times n_2$ matrix $M$ and a non negative integer $r\leq n_1,n_2$, we denote by $M_r$ the diagonal matrix whose diagonal entries are the $r$ largest singular values of $M$, and by $M_{r,c}$ the diagonal matrix whose diagonal entries are the remaining singular values of $M$.

The following theorem concerning reconstruction with respect to rank one measurements corresponding to the elements of a random tight frame is the main result of this note. 
\begin{theorem}  \label{mainTh1}
Consider the above measurement process  with $m$ measurement vectors  $a_i$ which are the (transposed) elements  of a random tight frame multiplied by $\sqrt m$.  Let $r\leq n$ and suppose that
\begin{equation}\label{bound:m}
m\geq C_1nr.
\end{equation}
Then with probability at least $1 - 3\mathrm{e}^{-C_2 m}$ it holds that for any  $X \in \mathcal H_n$, 
any solution $X^\sharp$  to the above convex optimization problem 
with noisy measurements $b = \mathcal A (X)+w$, where $\| w \|_{\ell_q} \leq \eta$,
obeys
\begin{equation}\label{err:bound1}
\| X - X^\sharp \|_2 \leq \frac{D_1}{\sqrt r}\norm{X_{r,c}}_1+D_2\cdot\frac{\eta}{ m^{1/q}}.
\end{equation}
 Here 
$C_1,C_2, D_1, D_2$ denote  positive universal constants. (In particular, for $\eta=0$ and $X$ of rank at most $r$ one has exact reconstruction.) 
\end{theorem}
This result is of similar nature as the main results of \cite{KKRT}, where however different types of measurements were analyzed. The bound \eqref{bound:m}
on the number of measurements is optimal.
In the special case $r=1$ it implies that a vector $x \in \RR^n$ (or in $\CC^n$) can be reconstructed robustly from noisy phaseless measurements
$y_j = |\langle x, a_j \rangle|^2 + w_j$ via the PhaseLift approach, see \cite{castvo13,KRT,KKRT} for details.

The above result is shown via establishing a version of the null space property of the measurement map $\nA$, namely the  Frobenius-robust rank null space property with respect to $\ell_q$.
Let us recall from \cite{KKRT} this notion (in the Hermitian case) which serves as a useful  recovery criterion for the above  measurement  process (\ref{eq:MeasurementProcess}) via the  minimization problem (\ref{eqNNMinimization}). 

\begin{definition} 
For $q\geq 1$, we say that $\nA:\mathcal H_n\to\RR^m$ satisfies the Frobenius-robust rank null space property with respect to $\ell_q$ of order $r$ with constants $0<\rho<1$ and $\tau>0$ if for all $M\in\mathcal H_n$, the singular values of $M$ satisfy
$$
\norm{M_r}_2\leq\frac{\rho}{\sqrt r}\norm{M_{r,c}}_1+\tau\norm{\nA(M)}_{\ell_q}.
$$
\end{definition} 

In \cite[ Theorem 3.1 and Remark 3.2]{KKRT}, it is shown  for the above measurement process (\ref{eq:MeasurementProcess}) 
that if $\nA$ satisfies the Frobenius-robust rank null space property with respect to $\ell_q$ of order $r$ (with constants  $0<\rho<1$ and $\tau>0$)
then any solution $X^{\sharp}$ of (\ref{eqNNMinimization}) approximates $X$ with error $$\| X - X^\sharp \|_2 \leq \frac{C_1(\rho)}{\sqrt r}\norm{X_{r,c}}_1+ C_2(\rho){\tau \eta}.$$ Here $C_1(\rho)$ and $C_2(\rho)$ are explicit positive constants depending only on $\rho$. \newline


{\em Proof of Theorem~\ref{mainTh1}.}
Let $Q$ be the matrix whose rows  $q_i^*$ are the elements of our random tight frame multiplied by $\sqrt m$.
Recall that an  $n\times n$ Wishart matrix with $m$ degrees of freedom is a random matrix of the form $AA^*$, where
 $A$ is an $n\times m$ Gaussian matrix. Let $W$ be $1/\sqrt{m}$  times the square root of  an $n\times n$ Wishart matrix with $m$ degrees of freedom (i.e. $W=\frac{1}{\sqrt m}\sqrt{AA^*}$, where $A$ is as above), independent of $Q$. Then  $G:=QW$ is an $m\times n$ Gaussian matrix, cf. \cite[Proposition 9]{TroppComparison}. Let $g_1^*,\hdots, g_m^*$ be the rows of $G$.  Then $g_i=W^*q_i$. We denote by $\nA^{(G)}$  the above measurement map $\nA$   with measurement  vectors $a_i=g_i$.   Similarly, we denote by $\nA^{(Q)}$ be the above map  $\nA$ with measurement  vectors $a_i=q_i$. It follows that for any Hermitian $n\times n$ matrix $X$  $$\nA^{(G)}(X)_i=\tr(Xg_ig_i^*)=\tr(XW^*q_iq_i^*W)=\tr(WXWq_iq_i^*)=\nA^{(Q)}(WXW)_i.$$
Hence $$\nA^{(Q)}(X)=\nA^{(G)}(W^{-1}XW^{-1})$$ for all $X$.

 Now \cite[Lemma 8.1]{KKRT} tells us that if $\nA^{(G)}$  satifies the Frobenius robust rank null space property with respect to $\ell_q$ of order $r$ with constants $\rho, \tau$, and if $ \kappa(W)^{2}\rho<1$, then the map $X\mapsto \nA^{(G)}(W^{-1}XW^{-1})$ satisfies the  Frobenius robust rank null space property with respect to $\ell_q$ of order $r$ with constants $\kappa(W)^{2}\rho, \norm{W}_{\infty}^{2}\tau$. (Here $\kappa(W)=\frac{\lambda_{\max}(W)}{\lambda_{\min}(W)}$ is the quotient of the maximal and the minimal eigenvalue of $W$.)
 Assuming that $\nA^{(G)}$ satifies the Frobenius robust rank null space property with respect to $\ell_q$ of order $r$ with constants $\rho, \tau$ such that  $\kappa(W)^{2}\rho<1$,  it follows that $\nA^{(Q)}$  satisfies the   Frobenius robust rank null space property with respect to $\ell_q$ of order $r$ with constants $\kappa(W)^{2}\rho, \norm{W}_{\infty}^{2}\tau$. 

It remains to show that with high probability  $\nA^{(G)}$ satifies the Frobenius robust rank null space property with respect to $\ell_q$ with suitable constants    $0<\rho<1$ and $\tau>0$ such that  $\kappa(W)^{2}\rho<1$.

It is shown in the proof of \cite[Proposition 8.1]{KKRT}  that for any $c_1>1$ there are positive constants $c_2,c_3$ such that for $m\geq c_2n$ with probability at least $1-2e^{-c_3m}$ we have $$
 \max(\kappa(W), \norm{W}_{\infty})\leq c_1.
$$
Fix now $\rho=1/2$. Then, as shown in the proof of
 \cite[Theorem 1.2]{KKRT}, there are  positive constants $c_4,c_5, c_6$ such that,  for $m\geq c_4rn$,  with probabilty at least $1-e^{-c_5m}$, the map $\nA^{(G)}$ satisfies the Frobenius robust rank null space property with respect to $\ell_q$ of order $r$ with constants $\rho=1/2, \tau=c_6/m^{1/q}$. (This result makes heavy use of Mendelson's small ball method, \cite{KoltchinskiiMendelson,Mendelson,tr14}.) Choose now $1<c_1<\sqrt{2}.$
Then with probability at least  $1-2e^{-c_3m}$ we have $ \kappa(W)^{2}\rho\leq c_1^2/2<1$ and $\norm{W}_{\infty}^2\leq c_1^2< 2.$ It follows that for $m$ large enough (as above) with probability at least $1-3e^{-C_2m}$, the map $\nA^{(Q)}$  satisfies the Frobenius robust rank null space property with respect to $\ell_q$ of order $r$ with constants $ \kappa(W)^{2}/2<1$ and $2c_6/m^{1/q}$. Hence by the above mentioned recovery result \cite[ Theorem 3.1]{KKRT}, for $m\geq C_1nr$, with probability at least $1-3e^{-C_2m}$ the recovery result of the Theorem holds (where $C_1=c_4$ and $C_2=\min\{c_3,c_5\}$ are universal positive constants). \hfill $\square$

\begin{remark}
The multiplication of  the $q_i$ by $\sqrt m$ yields the correct normalization compared to Gaussian measurements since the sum of the squared $\ell_2$-norms of all $m$ elements in the random tight frame is $n$, hence  on average each row vector has squared $\ell_2$-norm equal to $n/m$, whereas for  Gaussian measurement vectors the  squared    $\ell_2$-norm has expectation $n$.
\end{remark}

\noindent {\em Acknowledgement.} We thank Richard K\"ung for helpful discussions.

\end{document}